\documentclass[aps,preprint,showpacs,groupedaddress]{revtex4}
\usepackage{graphicx} 
\usepackage{amssymb} 
\usepackage{amsmath}
\usepackage{bbold}
\usepackage{dsfont}
\usepackage{color} 
\usepackage[pstricks,mediumqspace,squaren]{SIunits} 
\newcommand{\defined}{\ensuremath{\mathrel{:=}}}
\newcommand{\Teflon}{Teflon\textsuperscript\textregistered~}

\begin{document}
\title{{Rabi~Oscillations~at~Exceptional~Points~in~Microwave~Billiards}}
\author{%
  B.~Dietz,$^1$ %
  T.~Friedrich,$^1$ %
  J.~Metz,$^1$ %
  M.~Miski-Oglu,$^1$ %
  A.~Richter,$^1$\footnotemark[1]\footnotetext[1]{richter@ikp.tu-darmstadt.de} %
  F.~Sch{\"a}fer,$^1$ %
  C.~A.~Stafford.$^2$ %
} %
\affiliation{%
  \centerline{%
    $^1$Institut~f{\"u}r~Kernphysik, %
    Technische Universit{\"a}t~Darmstadt, %
    D-64289~Darmstadt, %
    Germany} %
  \centerline{%
    $^2$Physics~Department, %
    University~of~Arizona, %
    1118~East~4th~Street, %
    Tucson,~Arizona~85721} %
} %
\begin{abstract}
  We experimentally investigated the decay behavior with time $t$ of resonances near and at
  exceptional points, where two complex eigenvalues and also the associated 
  eigenfunctions coalesce. The measurements were performed with a dissipative microwave billiard, 
  whose shape depends on two parameters.
  The $t^2$-dependence predicted at the exceptional point on the basis of a 
  two-state matrix model could be verified.
  Outside the exceptional point the predicted Rabi oscillations, also called quantum echoes 
  in this context, were detected. 
  To our knowledge this is the first time that 
  quantum echoes related to exceptional points were observed experimentally.
\end{abstract}

\maketitle

In quantum mechanics many dynamical processes are dominated by (avoided) level 
crossings. A crossing of two eigenvalues requires the variation of at least two 
parameters. It has been known for many years, that near a crossing the two energy surfaces 
form two sheets of a double cone~\cite{Neumann:29,Teller:37}. The
apex 
of the double cone is associated with a singularity and called diabolic
point~(DP)~\cite{Berry:84-DP,Berry:84}. A DP occurs in
Hermitian Hamiltonians. Phenomena related to
a DP, as e.g.\ geometric phases, have been studied theoretically
in various generalizations of Berry's original paper (see e.g.\ \cite{Shapere:89,Anandan:97}
and references therein) and experimentally e.g.\ in~\cite{Lauber:94}~with a microwave billiard.
For non-Hermitian Hamiltonians, 
as those used for the theoretical description of dissipative systems, a topologically
different singularity may appear: an exceptional point (EP)~\cite{Kato:66} 
-- there not only the eigenvalues but also the associated eigenstates 
coalesce~\cite{Harney:00}. 
Thus EPs are not only singularities of the
spectrum but also of the eigenstates. They have been observed in
laser induced ionizations of atoms~\cite{Latinne95}, %
crystals of light~\cite{Oberthaler96}, %
electronic circuits~\cite{Stehmann04}, %
the propagation of light in dissipative media~\cite{Shuvalov00,BerryEP:03} %
and in microwave billiards~\cite{Dembo01,Dembo03,Dembo04} %
and also appear in many theoretical models: e.g.\ in that used for 
the decay of superdeformed
nuclei~\cite{echos_theorie}, phase transitions and avoided
level crossings~\cite{Heiss91,Keck:03}, geomagnetic polarity
reversal~\cite{Stefani:05}, tunneling between quantum
dots~\cite{cardamone2002}, and in the context with the crossing of two Coulomb
blockade resonances \cite{Weiden:03}.

Exceptional points give rise to interesting phenomena such as level crossings 
and
geometric phases~\cite{Heiss90,Berry:84,Dembo04}. In this article we present
novel experimental results on the time decay of resonances in the vicinity of and at
an EP. Close to an EP, the time spectrum exhibits --
beside the decay of the isolated resonances -- oscillations with a fixed
frequency; these are called quantum echoes \cite{Thomas:04}. Quantum echoes 
occur due to the
transfer of energy between the two nearly degenerate resonances.  
At the EP their vanishing
and a quadratic time dependence of the resonance amplitude was predicted. This time
dependence is a characteristic property of EPs with exactly two
coinciding eigenvalues. In general
  more than two eigenvalues may coincide at an EP, however 
  the coincidence of two eigenvalues is the most probable. 

\begin{figure} %
  \centering %
  \includegraphics[width=0.95\columnwidth]{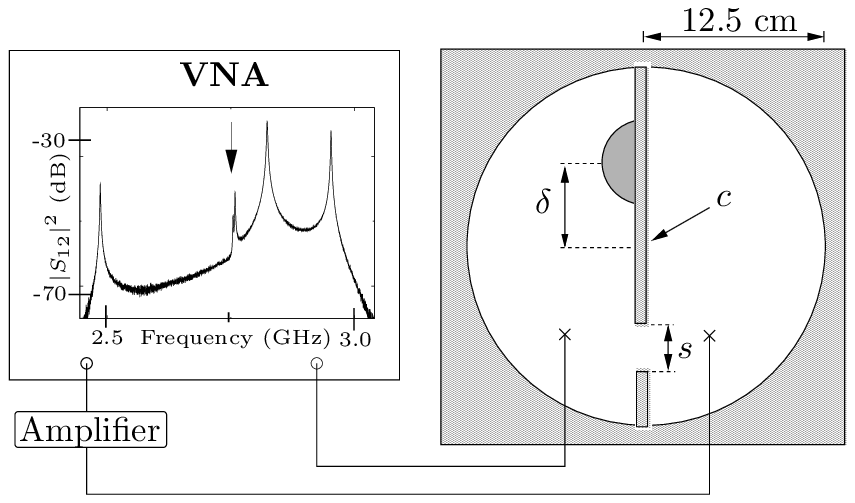}
  \caption{%
    Sketch of the experimental setup: In the microwave billiard (right part) 
    the parameter 
    $s$ refers to the length
    of the slit and $\delta$ to the position of the semi-circular 
    \Teflon disc,
    measured from the billiard center $c$. 
    The antenna positions are shown as crosses.
    The vectorial network analyzer 
    (left part) measures the ratio of the power of the signals received at 
    the left antenna $1$ and
    emitted at the right antenna $2$, $\vert S_{12}\vert^2$, and their relative 
    phase. Shown is a part of a typical transmission spectrum 
    measured in dB. The arrow points at two nearly degenerate resonances which
for a certain choice of the parameters $s$ and $\delta$ 
    coalesce at the exceptional 
    point at $\unit{2.757\pm 0.001}{GHz}$.
}
  \label{fig:MWBskizze}
\end{figure}

For wavelengths longer than twice the height of the microwave billiard the
scalar Helmholtz equation of the electric field strength in a cylindric
microwave billiard is equivalent to the Schr{\"o}dinger equation for the wave
functions in a quantum billiard of corresponding shape (see
e.g.\ ~\cite{Stoeckmann:buch00,richter97}). Thus, aside from their intrinsic interest,
flat microwave billiards yield a possibility to gain experimental insight into
properties of the eigenvalues and eigenfunctions of two-dimensional quantum
billiards. We used a coupled pair of dissipative, cylindric and flat microwave 
billiards manufactured of copper to mimic the two-level system required for an
EP to occur~\cite{Dembo01,Dembo03,Dembo04}; a sketch is shown
in Fig.~\ref{fig:MWBskizze}. It is obtained by
dividing a circular microwave billiard with a barrier made from copper into 
two approximately equal parts differing in their areas by about 5\%.
Due to this slight difference in size the
twofold degeneracy of the eigenfrequencies of the circular billiard is lifted,
that is the eigenfrequencies are split into two
nearly degenerate ones. 
Their crossing behavior has been analyzed in~\cite{Dembo01,Dembo03,Dembo04}
with a slightly modified experimental setup. 
In order to control the coalescence of two states, we must be able to
control their frequencies and widths. We achieved this by
tuning two parameters, the length of slit $s$ between the two parts of
the microwave billiard and the position $\delta$ of a semicircular
\Teflon disc (see Fig.~\ref{fig:MWBskizze}). The parameter $\delta$ 
mainly affects the resonance frequencies of the billiard part
which contains the \Teflon disc, whereas
the parameter $s$ controls the coupling between
the eigenmodes of the two billiard parts. Coupling zero is achieved by closing the
slit. Then, due to the Dirichlet boundary condition, the electric field strength vanishes 
alongside the barrier. When opening the slit, those eigenmodes of the resonators are
affected, which have a non vanishing field intensity near the slit. Depending on their derivatives
in normal direction to the slit, they penetrate into the other cavity
and couple to its modes.

To measure the transmission spectrum of the billiard system the vectorial 
network analyzer
``HP-8510C'' (VNA) produces an rf signal, which is piped trough a
\unit{30}{dB} amplifier and is forwarded to a dipole antenna with
\unit{0.5}{mm} in diameter and \unit{2}{mm} in length. If the signal frequency is
tuned to a resonance frequency of the billiard, the dipole antenna
excites the corresponding eigenmode such that power is radiated into the
microwave billiard -- otherwise the power is reflected at the antenna.  
The power is coupled out via a second antenna and lead back into the 
VNA, which compares the produced and the received signal in amplitude and
phase. 
The squared eigenfunctions are measured with the perturbation body method.  
This method is
based on the Maier-Slater theorem~\cite{slatertheorem}, which states that the
amount of the frequency shift of a resonance depends on the
difference of the square of the electric field strength and of the
magnetic induction at the position of the perturbing body. 
Hence, the intensity distribution of the electric field strength is obtained 
by moving the perturbation body through the microwave billiard on a grid with 
spacing \unit{5}{mm} and comparing the perturbed and unperturbed resonance frequency.
As perturbation we use a cylindrical magnetic rubber with about \unit{2.2}{mm} 
in diameter and \unit{3.5}{mm} in height.
Magnetic rubber has the advantageous property, that it suppresses the 
contribution of the magnetic induction to the frequency shift
and still enables us to change its position inside the
closed microwave billiard with a guide magnet. 

Quantum echoes are time recurrent signals. They have been seen e.g.\ in open
microwave billiards \cite{Thomas:04} but they also appear in the gravity induced
interference effect~\cite{Colella}, the Aharonov-Bohm effect~\cite{Aharonov:59}
as well as in $\textrm{NH}_3$--MASERs ~\cite{Gordon}. In the latter they are better
known as a beating with a frequency called Rabi frequency. 
The decay behavior with time of two resonance states which are near an EP, 
i.e.\ the superposition of the energy transfer between the two resonances
and that to the exterior, is analyzed with a two-level Hamiltonian 
which was used
by Stafford and Barrett~\cite{echos_theorie} in a different context.
The theoretical results presented there may be mapped one to one onto our problem.
This signifies, that the experimental results presented here are of general interest in different
fields of physics dealing with EPs.

The electric field strengths $\vec {\mathbb E}_j(t) = E_j(t)\, \vec e_z$ with
$j=1,2$ of the eigenmodes in the two parts of the microwave billiard can be
described as a pair of coupled damped harmonic oscillators,
\begin{align}
  \label{eq:harm-osci}
  \left( %
    \frac{\partial^2}{\partial t^2} 
    + \gamma_1 \frac{\partial}{\partial t}
    + \omega_1^2
  \right) %
  E_1(t) &= F_1(t) - \sigma E_2(t)  \nonumber \\
  \left( %
    \frac{\partial^2}{\partial t^2} 
    + \gamma_2 \frac{\partial}{\partial t}
    + \omega_2^2
  \right) %
  E_2(t) &= F_2(t) - \sigma E_1(t),
\end{align}
with angular frequencies $\omega_1$ and $\omega_2$, respectively, decay
rates $\gamma_1$ and $\gamma_2$ and a coupling strength $\sigma $. 
The driving terms $F_j(t)$ with $j=1,2$
describe the coupling of the antennae to the eigenmodes.  
The coupled differential equations~(\ref{eq:harm-osci}) can be solved with the
Green function method. The inverse of Green's function is given
by $\mathbb{G}^{-1}(\omega) =  \omega^2\mathbb{1} - \mathbb{H}(\omega)$ 
with the non-Hermitian two-level Hamiltonian \cite{Dembo03,Dembo04}
\begin{equation}
    \label{eq:Echo-Hamilton}
    \mathbb{H}(\omega) = 
    \left( %
      \begin{array}{cc} %
        \omega_1^2 - i \, \omega \, \gamma_1 & \sigma \\ %
        \sigma & \omega_2^2 - i\, \omega \, \gamma_2 %
      \end{array} %
    \right).
\end{equation}
While the real parts of the diagonal elements ($\omega_1^2$ and $\omega_2^2$) are
proportional to the square of the  resonance frequencies of the uncoupled billiard
parts, the imaginary parts ($\Gamma_1 \equiv \omega \, \gamma_1$ and
$\Gamma_2 \equiv \omega \, \gamma_2$) describe their resonance widths.  The
Hamiltonian is symmetric since the system is time reversal invariant and it is
non-Hermitian due to the antennae and the absorption in the billiard walls and
the \Teflon disc. We denote the eigenvalues of Eq.~(\ref{eq:Echo-Hamilton}) by
${\cal E}_\pm \defined \bar \omega^2 - i \bar \Gamma \pm R$ with %
$R\defined \frac{1}{2}\sqrt{[\Delta \omega^2 - i \Delta \Gamma]^2 + 4\sigma^2}$, %
$\bar \omega\defined \sqrt{\left(\omega^2_1 + \omega^2_2 \right)/2}$, %
$\bar \Gamma \defined (\Gamma_1 + \Gamma_2)/2$, %
$\Delta \omega^2 \defined \omega_2^2 - \omega_1^2$ and %
$\Delta \Gamma \defined \Gamma_2 - \Gamma_1$. %
Although in the experiment the physical quantities $\omega_1$, $\omega_2$,
$\Gamma_1$ and $\Gamma_2$ are complicated functions of the \Teflon disc position
$\delta$ and the opening length $s$ (see Fig.~\ref{fig:MWBskizze}), these two parameters
are sufficient to achieve 
coalescence of the two eigenvalues of the two-level Hamiltonian, i.e.\ the
vanishing of the parameter $R$. At the EP $\Delta\omega^2$ vanishes and 
$\Delta\Gamma^2=4\sigma^2$. The parameter setting is denoted by $s=s^{\textrm{EP}}$ and 
$\delta =\delta^{\textrm{EP}}$.
As we are interested in the decay behavior with time of the two-level system we need
to compute the Fourier transformation from the frequency to the time domain of 
$\mathbb{G}(\omega)$. The resulting 
expression is quite complicated. However, in the parameter range of the
experiment a simple and accurate analytic expression is obtained with the 
following approximation: 
in the vicinity of $\omega \simeq \omega_1, \omega_2$ the terms $\omega \, \gamma_1$
and $\omega \, \gamma_2$ are small in comparison with $\omega^2$. Therefore
$\omega$ in Eq.~(\ref{eq:Echo-Hamilton}) is replaced by the average value of
the angular frequencies $\bar \omega$.  This approximation takes into account
that the major contribution comes from the poles of the Green function.
With
\begin{subequations} %
  \begin{align} %
    \Omega &\defined \frac{R}{2\; \bar \omega}~, & %
    f &\defined \bar \omega - \frac{i}{2} \; \bar \gamma~, & %
    \bar \gamma \defined \frac{\gamma_1 + \gamma_2}{2} %
    \label{eq:echo_frequenz}
  \end{align} %
  the Fourier transform $\mathbb{\tilde G}_{12}(t)$, which describes the
  energy transfer from one resonance to another, is given
  by~\cite{echos_theorie,cardamone2002}
  \begin{equation} %
    \left| %
      \mathbb{\tilde G}_{12}(t)  %
    \right|^2 %
    \approx %
    \frac{\sigma^2}{\bar \omega^2} %
      \left| %
        \frac{\textrm{e}^{- i f t}}{\sqrt{{\cal E}_{+} {\cal E}_{-}}} \; %
        \big[ %
        \cos{(\Omega \, t)} %
        + i f \; \frac{\sin{(\Omega \, t)}}{\Omega} %
        \big] %
      \right |^2 %
    \label{eq:echo-Formel}. %
  \end{equation} %
\end{subequations} %
Hence, $\mathbb{\tilde G}_{12}(t)$ decays exponentially with a decay constant of $\bar \gamma/2$ 
and oscillates with
a high, not resolvable angular frequency $\bar \omega$. The physical value
$\Omega/2\pi$ denotes the much lower ($R \ll \bar \omega$) \emph{echo
  frequency}. It corresponds to the Rabi frequency first observed in NMR and optical
pumping \cite{Rabi} and also
well known in quantum optics~\cite{Scully:97}, nuclear physics~\cite{echos_theorie} 
and quantum computing~\cite{Benenti:04}.

As was pointed out already in~\cite{echos_theorie,cardamone2002} the imaginary
part of the echo frequency $\Omega/2\pi$ adds to the decay, while the real
part describes the oscillation between the two resonances. As a consequence,
the quantum echoes vanish if $\Omega$ is purely imaginary. This happens
\emph{for all} subcritical couplings, that is for slit openings 
$s<s^{\textrm{EP}}$ at the critical \Teflon disc
position $\delta = \delta^{\textrm{EP}}$. For overcritical couplings, 
i.e.\ for $s>s^{\textrm{EP}}$, 
the quantum echoes persist for all \Teflon disc positions, while exactly
at an EP both the real and imaginary part of the echo frequency
$\Omega/2\pi$ vanish. Then, with Eq.~(\ref{eq:echo-Formel}) a quadratic time
dependency of the echo amplitude
\begin{equation}
  \label{eq:EP-Echoes}
  \lim_{\left(s,\delta\right)\rightarrow {\bf EP}} %
  \left| %
    \mathbb{\tilde G}_{12}(t) %
  \right|^2 %
  \approx \frac{\sigma^2}{{\cal E}_+{\cal E}_-} \; t^2 \;\textrm{e}^{-\bar \gamma \, t} %
\end{equation}
is obtained. This dependency can also be verified by an exact calculation
based on Eq.~(\ref{eq:Echo-Hamilton}) evaluated at the EP. It is
consistent with the result given in~\cite{Kirillov04}.

The EPs were localized in the parameter plane $\{s, \delta\}$ by
varying the position of the \Teflon disc and the slit opening and
measuring crossings and avoided crossings of the frequencies and widths of the
resonances which coincide at the EP~\cite{Brentano:00}. 
In parallel we performed measurements of the nodal domains of the 
eigenfunction. For a more detailed description see for example~\cite{Dembo04}. 
Since the time behavior is extremely sensitive in the vicinity of an EP, the
\Teflon disc position and the opening of the slit were changed in steps of 
\unit{0.5}{mm}. We were able to
localize two EPs below \unit{3}{GHz}, the first at $\unit{2.757\pm 0.001}{GHz}$ 
and the 
second at $\unit{2.806\pm 0.001}{GHz}$. In this frequency range the pairs of
resonances which coalesce at the EP can be treated as isolated. 
Hence the two-level model is applicable. Since both EPs exhibit the 
same decay behavior with time, we only show the results for the first.

\begin{figure} %
  \centering %
  \hspace*{-2ex} %
  \includegraphics[width=\columnwidth]{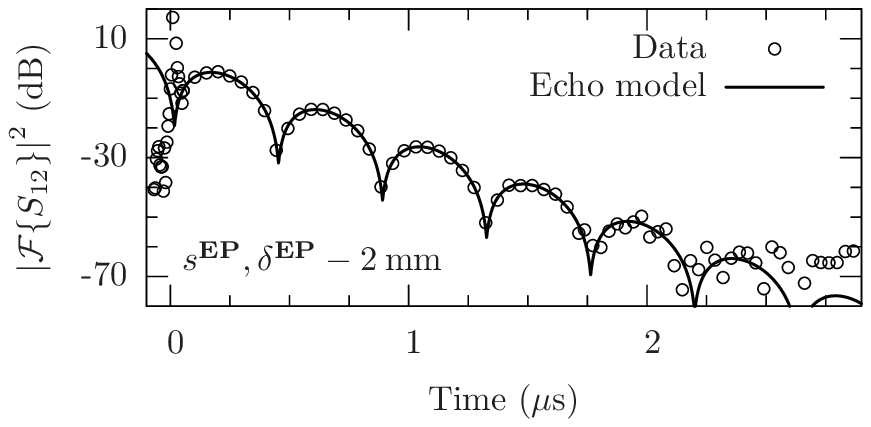}
  \caption{%
    The model (Eq.~(\ref{eq:echo-Formel})) describes the time spectrum very
    well. This is shown here for the exceptional point at  
    $\unit{2.757\pm 0.001}{GHz}$ for the parameter setting %
    $s = s^{\textrm{EP}} \approx \unit{48}{mm}$, %
    $\delta = \delta^{\textrm{EP}} - \unit{2}{mm} \approx \unit{21.5}{mm}$. %
    The echo frequency is $\Omega/2\pi \approx \unit{1.1\pm 0.2}{MHz}$. %
  }%
  \label{fig:anpassung-echoformel} %
\end{figure}
The decay of the resonances with time was deduced from the transmission
spectra by a fast Fourier transformation. For its computation a narrow frequency range of
about 0.3~GHz was chosen around the doublet and a Hamming window function was 
used. We checked that the choice of the latter does not affect the results. 
The square of this transformation and a fit of Eq.~(\ref{eq:echo-Formel}) to these data are shown in
Fig.~\ref{fig:anpassung-echoformel} for the critical coupling, that is for
$s=s^{\textrm{EP}}$
and a subcritical \Teflon disc position $\delta < \delta^{\textrm{EP}}$.  The model
(Eq.~(\ref{eq:echo-Formel})) describes the measured signal very well starting
from a time larger than~\unit{30 \pm 5}{ns}, where a peak indicates the time
the signal needs to travel through the coaxial cables connecting the
VNA with the antenna, up to times, where the noise level is
reached (at about \unit{-65}{dB}).  The echo frequency equals $\Omega/2\pi =
\unit{1.1 \pm 0.2}{MHz}$, which approximately corresponds to the spacing
between the two eigenfrequencies.

\begin{figure} %
  \centering %
  \hspace*{-2ex} %
\includegraphics[width=\columnwidth]{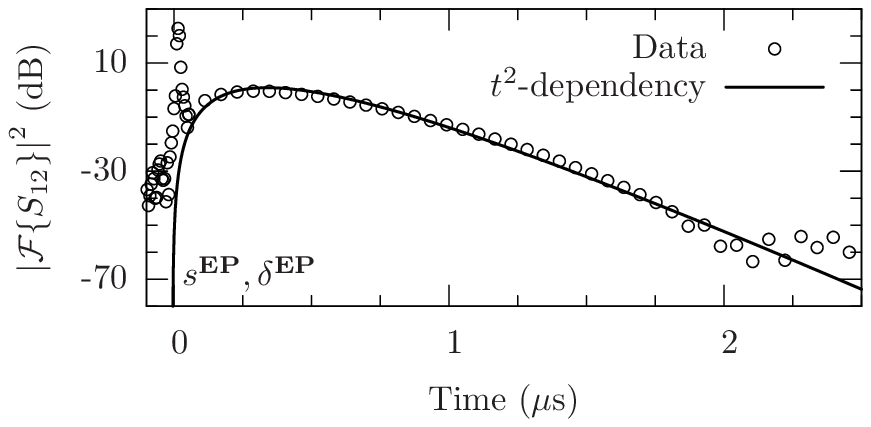}
  \caption{%
    At the EP the quantum echoes disappear and the predicted
    quadratic time decay of the echo amplitude could be verified. Shown is the
result for the EP at $\unit{2.757\pm 0.001}{GHz}$ together with the theoretical
prediction (Eq.~(\ref{eq:EP-Echoes})).} %
  \label{fig:EP1-t2-abhaengigkeit} %
\end{figure} %
The time decay of the resonances for $s=s^{\textrm{EP}}$
and $\delta = \delta^{\textrm{EP}}$ is shown in
Fig.~\ref{fig:EP1-t2-abhaengigkeit} together with the theoretical 
prediction 
(Eq.~(\ref{eq:EP-Echoes})). The quantum echoes disappear as they are
an interference effect of two eigenmodes. 
Moreover, the coalescence of the two eigenmodes has an
influence on the time decay behavior. An isolated resonance decays simply
exponentially, whereas at an EP the time dependence of the
echo amplitude is quadratic. 
Hence, the time decay behavior of the resonances provides 
information on the nature of a degeneracy, and, consequently, the location of EPs in the 
parameter plane. In order to check how close we are to the EP we performed an 
additional fit of the function in Eq.~(\ref{eq:EP-Echoes}) to the experimental 
data, however this time with a $t^\alpha$- instead of a $t^2$-dependence of the 
echo amplitude, yielding $\alpha = 1.91 \pm 0.04$. 
This small deviation from the $t^2$-dependence and that observed 
in Fig.~\ref{fig:EP1-t2-abhaengigkeit} for times larger than 
$t \approx \unit{2}{\mu s}$ are due to the fact that the
EP is not hit exactly. It expresses how sensitive this
experiment is to tiny changes in the parameters.  
If the real parts of the two eigenvalues of the Hamiltonian 
(Eq.~(\ref{eq:Echo-Hamilton})) 
coincide, $R$ and therefore $\Omega$ are purely imaginary for subcritical 
coupling, where $s<s^{\textrm{EP}}$, and we expect for the time dependence a 
sum of exponentials. 
Equation~(\ref{eq:echo-Formel}) again provides a very good description
of the measured time dependence. However, we were not able to verify that at the critical
Teflon position $\delta=\delta^{\textrm{EP}}$ the quantum echoes vanish for
subcritical couplings. It seems that the resonance frequencies 
$\omega_1$ and $\omega_2$ of the two uncoupled resonators are only approximately
degenerate, such that the real
part of $\Omega$ in Eq.~(\ref{eq:echo-Formel}) is very small but non vanishing.

In the present work we experimentally studied quantum echoes in the vicinity
of and at two exceptional points (EPs).  Although the system is very
sensitive to small deviations of the critical parameters, we were able to
control the system sufficiently well to verify the predicted $t^2$-dependence
of the echo amplitude at an EP. We showed, that a
$2\times2$-model describes the quantum echoes very well. To our knowledge
this is the first experimental verification of quantum echoes in the vicinity
of an EP. In addition, the determination of the time dependence of the transfer
of energy between two closely lying resonances 
provides a straightforward method for the location of exceptional points in the parameter
plane.

We thank O.~Kirillov for several discussions. Part of the theoretical work was carried 
out during a gathering at the Centro Internacional de Ciencias (CIC) in
Cuernavaca 2004. This work has been supported by the DFG within the SFB 634.
C.S. was supported by NSF grant No. 0312028.


\begin{thebibliography}{99}

\bibitem{Neumann:29} J. Von Neumann and E. P. Wigner, Z. Phys. {\bf 30}, 467 (1929). 
\bibitem{Teller:37} E. Teller, J. Phys. Chem. {\bf 41}, 109 (1937).
\bibitem{Berry:84-DP} M. V. Berry and M. Wilkinson, Proc. Roy. Soc. Lond. {\bf A392}, 15 (1984).
\bibitem{Berry:84} M. V. Berry, Proc. Roy. Soc. Lond. {\bf A392}, 45 (1984).
\bibitem{Shapere:89} A. Shapere and F. Wilczek, 
\emph{Geometric Phases in Physics} (World Scientific, Singapore, 1989).
\bibitem{Anandan:97} J. Anandan, J. Christian and K. Wanelik, Am. J. Phys. {\bf 65}, 180 (1997).
\bibitem{Lauber:94} H.-M. Lauber, P. Weidenhammer and D. Dubbers, Phys. Rev. Lett. {\bf 72}, 1004 (1994).
\bibitem{Kato:66} T. Kato, \emph{Perturbation theory of linear operators} (Springer,Berlin,1966).
\bibitem{Harney:00} H. L. Harney and W. D. Heiss, Eur. Phys. J. D {\bf 17}, 149
(2001).
\bibitem{Latinne95} O. Latinne {\it et al.}, Phys. Rev. Lett. {\bf 74}, 46 (1995).
\bibitem{Oberthaler96} M. K. Oberthaler {\it et al.}, Phys. Rev. Lett. {\bf 77}, 
4980 (1996).
\bibitem{Stehmann04} T. Stehmann, W. D. Heiss and F. G. Scholtz, J. Phys. A: Math. Gen. {\bf 37}, 7813 (2004).
\bibitem{Shuvalov00} A. L. Shuvalov, N. H. Scott, Acta Mech. {\bf 140}, 1 (2000).
\bibitem{BerryEP:03} M. V. Berry and M. R. Dennis, Proc. Roy. Soc. Lond. {\bf A459}, 
1261 (2003).
\bibitem{Dembo01} C. Dembowski {\it et al.}, Phys. Rev. Lett. {\bf 86}, 
787 (2001).
\bibitem{Dembo03} C. Dembowski {\it et al.}, Phys. Rev. Lett. {\bf 90}, 034101 (2003).
\bibitem{Dembo04} C. Dembowski {\it et al.}, Phys. Rev. E {\bf 69}, 056216 (2004). 
\bibitem{echos_theorie} C. A. Stafford and B. R. Barrett, Phys. Rev. C {\bf 60}, 051305 (1999).
\bibitem{Heiss91} W. D. Heiss and A. L. Sannino, Phys. Rev. A {\bf 43}, 4159 (1991).
\bibitem{Keck:03} F. Keck, H. J. Korsch and S. Mossmann, J. Phys. A: Math. Gen. {\bf 36}, 2125 (2003).
\bibitem{Stefani:05} F. Stefani and G. Gerbeth, Phys. Rev. Lett. {\bf 94}, 184506 (2005).
\bibitem{cardamone2002} D. M. Cardamone, C. A. Stafford and B. R. Barrett, Phys. Stat. Sol. B {\bf 230}, 419 (2002). 
\bibitem{Weiden:03} H.~A.~Weidenm\"uller, Phys. Rev. B {\bf 68}, 125326 (2003). 
\bibitem{Heiss90} W. D. Heiss and A. L. Sannino, J. Phys. A: Math. Gen. {\bf 23}, 1167 (1990).
\bibitem{Thomas:04}  C. Dembowski {\it et al.}, Phys. Rev. Lett. {\bf 93}, 134102 (2004).
\bibitem{Stoeckmann:buch00} H.-J. St{\"o}ckmann \emph{Quantum  Chaos: An Introduction}
(Cambridge University Press,Cambridge,2000).
\bibitem{richter97}  A. Richter, in: \emph{Emerging
Applications of Number Theory, The IMA Volumes in Mathematics and its
Applications}, Vol. {\bf 109}, edited by D.A. Hejhal
{\it et al.}, p. 479, (Springer, New York, 1999).
\bibitem{slatertheorem} L.C. Maier and J.C. Slater, J. Appl. Phys. {\bf 23}, 68 (1952).
\bibitem{Colella} J.-L. Staudenmann {\it et al.} Phys. Rev. A {\bf 21}, 1419 (1980).
\bibitem{Aharonov:59} Y.~Aharonov and D.~Bohm, Phys. Rev. {\bf 115}, 485 (1959); Y.~Aharonov
and J.~Anandan, Phys. Rev. Lett. {\bf 58}, 1593 (1987). 
\bibitem{Gordon} J.~P.~Gordon, H.~J.~Zeiger, and C.~H.~Townes, Phys. Rev. {\bf 99}, 1264 (1955).
\bibitem{Rabi} I.~I.~Rabi {\it et al.} Phys. Rev. {\bf 55}, 526 (1939); W.~Happer, Rev. Mod. Phys. 
{\bf 44}, 169 (1972).
\bibitem{Scully:97} M. O. Scully and M. S. Zubairy, \emph{Quantum Optics} (Cambridge University Press,Cambridge,1997).
\bibitem{Benenti:04} G. Benenti, G. Strini and G. Casati, \emph{Principles of Quantum Computation and Information} (World Scientific,Singapore,2004).
\bibitem{Kirillov04} O. N. Kirillov and A. P. Seyranian, SIAM J. Appl. Math. {\bf 64}, 1383 (2004). 
\bibitem{Brentano:00} M. Phillipp, P. von Brentano and G. Pascovici and A. Richter, Phys. Rev. E {\bf 62}, 1922 (2000). 
\end{thebibliography}
\end{document}